\documentclass[epsfig]{aa}
\usepackage{epsfig,deluxe}
\begin{document}

\newcommand{\gsim}{\hbox{\rlap{$^>$}$_\sim$}}
  \thesaurus{06;  19.63.1}
%
\authorrunning{A. Dar \& A. De R\'ujula}
\titlerunning{A CB model of GRBs: superluminal signatures}
\title{A cannonball model of gamma-ray bursts: superluminal
signatures} 

\author{Arnon Dar$^{1,2}$ and A. De R\'ujula$^1$}
\institute{1. Theory Division, CERN, CH-1211 Geneva 23, Switzerland\\ 
           2. Physics Department and Space Research Institute, Technion,
              Haifa 32000, Israel } 
\maketitle

\begin{abstract} 

Recent observations suggest that the long-duration gamma ray bursts (GRBs)
and their afterglows are produced by highly relativistic jets emitted in
supernova explosions. We propose that the result of the event is not just
a compact object plus the ejecta: within a day, a fraction of the parent
star falls back to produce a thick accretion disk. The subsequent accretion
generates jets and constitutes the GRB ``engine'', as in the observed
ejection of relativistic ``cannonballs'' of plasma by microquasars and
active galactic nuclei. The GRB is produced as the jetted cannonballs 
exit the supernova shell reheated by the collision, re-emitting their own 
radiation and boosting the light of the shell. They decelerate by
sweeping up interstellar matter, which is accelerated to cosmic-ray
energies and emits synchrotron radiation: the afterglow. We emphasize here
a smoking-gun signature of this  model of GRBs: the superluminal
motion of the afterglow, that can be searched for ---the sooner the better---
in the particular case of GRB 980425.

\end{abstract} 


\section{Introduction}

The rapid directional localization of gamma-ray bursts 
(GRBs) by the satellites
BeppoSAX (e.g. Costa et al. 1997), Rossi (e.g. Levine et al. 1996) and
by the IPN (Inter-Planetary Network)  of spacecrafts
(e.g. Cline et al. 1999) led to the discovery
of long-duration GRB afterglows, to  determinations of their redshifts
that verified their cosmological origin (Meegan et al. 1992), to the
identification of their birthplaces ---mainly star formation regions in
normal galaxies--- and to evidence of a possible association 
of GRBs with supernova
explosions (Galama et al. 1998; Kulkarni et al. 1998a). The enormous 
isotropic energies inferred from
the redshifts and  fluences  and the short-time variability have 
 demonstrated that GRBs
are produced by highly collimated emissions from gravitational stellar
collapse (Shaviv and Dar 1995; Dar 1998a and references therein), but the specific progenitors and production mechanisms are unknown. 
In Table I we summarize some properties of the localized GRBs, for
a currently favoured $\Omega\!=\! 1$ Universe with 
cosmological constant  ${\rm \Omega_\Lambda\!=\! 0.7}$,
matter density ${\rm \Omega_M\!=\! 0.3}$ (all in critical units),
and Hubble constant ${\rm H_0\!=\! 65}$ km/(s Mpc).

The prevalent widespread belief is that GRBs are generated by synchrotron
emission from fireballs or firecones produced by mergers of compact stars 
(Paczynski 1986, Goodman et al. 1987) or by hypernova explosions 
(Paczynski 1998).
However, various observations suggest that most GRBs are produced in
supernova events by highly collimated superluminal jets (for more
details, see Dar and De R\'ujula 2000a).

In this paper, we outline a relativistic-cannonball model of GRBs,
which, we contend,
 explains the main observed features of GRBs and their afterglows.
We concentrate here on a 
major prediction of the model: the superluminal motion of
the GRBs afterglow.  This specific
signature, as we shall see, can still be searched for
---with a measure of urgency--- in the very particular case of
SN1998bw/GRB 980425.

\section{The supernova--GRB association}

There is mounting evidence for an association of supernova (SN)
explosions and GRBs. The first example was GRB 980425
(Soffita et al. 1998; Kippen 1998), within whose error circle
SN1998bw was soon detected optically  (Galama et al. 1998)
and at radio frequencies (Kulkarni et al. 1998a). The chance probability 
for a spatial and temporal coincidence is less than $10^{-4}$
(e.g. Galama et al. 1998), or much smaller if the revised BeppoSAX position
(e.g. Pian, 1999) is used in the estimate. The unusual radio 
(Kulkarni et al. 1998a; Wieringa et al. 1999)) and 
optical (Galama et al. 1998; Iwamoto et al. 1998) properties of SN1998bw, 
which may have been blended with the afterglow of GRB 980425,
support this association. The exceptionally small fluence
and redshift of GRB 980425 make this event very peculiar, a fact
that we discuss in detail below.

Evidence for a SN1998bw-like contribution to a GRB afterglow 
(Dar 1999a) was first found by Bloom et al. (1999) for GRB 980326, 
but the unknown redshift prevented a quantitative analysis.
The afterglow of GRB 970228 (located at redshift $\rm z=0.695$)
appears to be overtaken
by a light curve akin to that of SN1998bw (located at $\rm z_{bw}=0.0085$), 
when properly scaled by their differing redshifts (Dar 1999b).
Let the energy flux density  of SN1998bw be $\rm F_{bw}[\nu,t]$.
For a similar supernova located at z:
\begin{eqnarray}
{\rm F[\nu,t] = }&&{\rm{1+z \over 1+z_{bw}}\; 
{D_L^2(z_{bw})\over D_L^2(z)}}\, \times\nonumber \\
&&{\rm F_{bw}\left[\nu\,{1+z \over 1+z_{bw}},t\,
{1+z_{bw} \over 1+z}\right]\; A(\nu,z)}\, ,
\label{bw}
\end{eqnarray}
  
where $\rm A(\nu,z)$ is the extinction along
the line of sight.
The SN--GRB association in the case of GRB 970228 was 
reconfirmed by Reichart (1999) and by
Galama et al. (2000).  Evidence of  similar associations is found
for GRB 990712 (Hjorth et al. 1999; Sahu et al. 2000), GRB 980703 (Holland 
2000; Dar and De R\'ujula 2000a) and GRB 000418: an example that we show
in Fig.~1. In the case of GRB 990510, 
the evidence is marginal (Dar and De R\'ujula
2000a). For the remaining cases in Table I the data preclude a
conclusion, for one or more reasons: the late afterglow is not 
measured; $\rm F_{bw}[\nu']$ is not known for large
$\rm \nu'\simeq \nu\,(1+z)$; the GRB's afterglow or the host galaxy are
much brighter than the SN. The case of GRB 970508, for
which the afterglow is brighter than a SN contribution
given by Eq.~(\ref{bw}), is shown in Fig.~2. 

All in all, it is quite possible that a good fraction of GRBs be
associated with core-collapse 
SNe, perhaps even {\it all} of the most frequent,
long-duration GRBs.
The converse statement ---that most SNe of certain
types are associated with GRBs--- appears at first sight to be untenable.
The rate of Type Ib/Ic/II SNe has been estimated at ${\rm R_{SN}\sim
1/50~y^{-1}}$ per ${\rm L_*}$ galaxy (e.g. Van den Bergh \& Tammann 1991)
while the rate of GRBs (for our adopted cosmological parameters)
is a mere ${\rm R_{GRB} \sim 5\times 10^{-9}~y^{-1}}$ per ${\rm L_*}$
galaxy. Thus, very few of these SNe produce {\it visible} GRBs. But, if
the SN-associated GRBs were beamed within an angle $\theta\sim 10^{-3}$,
only a fraction $\pi\,\theta^2/4\pi \sim 2.5\times 10^{-7}$ would be
visible, making the observed rates compatible and making possible a 
rough one-to-one SN--GRB association
(or a ten-to-one association for $\theta\!\sim\! 3\times 10^{-3}$). 
Beaming within such narrow cones also solves the GRB ``energy crisis'' 
(Dar and Plaga 1999). 

To proceed, we need some understanding of the generation of superluminal
jets in SNe.  Lacking definite theoretical guidance, we turn to
observations in search for enlightment. 

\section{Superluminal jets}

Relativistic jets seem to be emitted by all astrophysical systems wherein
mass is accreted at a high rate from a disk onto a central compact object 
(for a recent review, see Mirabel and Rodriguez 1999a). 
Highly relativistic jets have been observed with high resolution in galactic
superluminal sources, such as the microquasars GRS 1915+105 (Mirabel and
Rodriguez 1994, 1999a,b;  Rodriguez and Mirabel 1999)
and GRO J165-40 (Tingay et al. 1995) where mass is accreted onto a stellar
black hole, and in many active galactic nuclei (AGNs) hosting a massive
black hole. These jets are not continuous streams: they consist of
individual blobs of plasma (plasmoids or ``cannonballs''). The firing of
these cannonballs (CBs) coincides with a sudden removal of the
accretion-disk material (Mirabel and Rodriguez 1999b). 

As they travel, microquasar CBs expand at the speed of a relativistic
plasma (c/$\sqrt{3}$ in their rest system) presumably due to the
pressure of their enclosed radiation. As they become transparent
and cool down, their lateral size stabilizes to a roughly constant
value, presumably constrained by magnetic
self-containment and/or by the ram pressure of the ambient material. Quasar
CBs show no measurable expansion as they travel, sometimes for as long as
Mpcs. Galactic and quasar CBs expand explosively when finally 
stopped by the
material they traverse.

Microquasars CBs are emitted in pairs, at similar velocities, ${\rm
v=\beta\, c}$, and angles $\theta$ and $\sim\!\pi\! -\!\theta$ relative to the line
of sight. Their apparent transverse velocities are 
$ {\rm v_T^\pm=\beta\,c\,\sin\theta/(1\mp\beta\,\cos\theta)}\, $.
For $\beta\,(\sin\theta+\cos\theta)\! >\!1$ the
approaching CB  is superluminal: ${\rm v_T^+ > c}$,
a case of ``relativistic aberration'' (Rees 1966).
For $\theta\ll 1$ and $\gamma\gg 1$: 
\begin{equation}
{\rm v_T^+\approx {2\,\gamma^2\,\theta\over (1+\gamma^2\theta^2)}\;c}\, ,
\label{superlum}
\end{equation}   
yielding  superluminal velocities ${\rm v_T\approx 2\,c/\theta}$
for $\gamma^2\theta^2\gg 1$, 
and ${\rm v_T\approx 2\,c\,\gamma^2\,\theta}$,
for $\gamma^2\theta^2\ll 1$, 
provided ${\rm 2\,\gamma^2\,\theta\! >\! 1}$.  
Superluminal velocities of a few c have been observed in microquasars, and
as high as 30 c in quasars (Gabudza et al. 1993).

\section{The cannonball model of GRBs}

The ejection of matter in a SN explosion is not fully understood. The
known mechanisms for imparting the required kinetic energy to the ejecta
are inefficient. Simulations of stellar collapse for parent stars rotating
at a typical observed rate give rise to proto-neutron stars spinning much
faster than young pulsars (Heger et al. 2000). Magnetic fields
no doubt play an important role in the outwards transport of angular
momentum, but current simulations (e.g. Spruit and Phinney, 1998) cannot fully
cope with their effects. In the related problem of accretion disks, a
freely parametrized bulk viscosity must be introduced to describe the
efficient angular momentum transport, presumably due to turbulent
magnetohydrodynamic effects (e.g. Longair, 1994). For all these reasons,
our theoretical understanding of core-collapse SN events is still unsatisfying.

We posit that the result of a SN event is not just a compact object plus
the ejecta: a fraction of the parent star falls back onto the newly born
compact object (De R\'ujula 1987). The free-fall time from a parent
stellar radius ${\rm R_\star}$ into an object of mass ${\rm M_c}$ is: 
\begin{eqnarray}
  {\rm t_{fall}}&&{\rm ={\pi\,\left[{R_\star^3\over 8\,G\,M_c}\right]^{1/2} }}
\nonumber \\
&&{\rm \sim 1\; day\; \left[{R_\star\over 10^{12}\;cm}\right]^{3/2}\;
    \left[{1.4\;M_\odot\over M_c}\right]^{1/2}}\, .
\label{tfall}
\end{eqnarray}
  
For a parent star that rotates, this time would be
longer, but for the material that falls from the polar regions.
It is quite natural to suppose that the infalling material settles
into an orbiting disk, or a thick torus if its mass is comparable
to ${\rm M_c}$. We assume that, as in other cases of significant accretion
into a compact object (microquasars and AGNs), jets of 
relativistic CBs of plasma are ejected. We presume their
composition to be ``baryonic'', as it is in the jets of
SS 433, from which Ly$_\alpha$ and Fe K$_\alpha$ lines have been detected
(Margon 1984), although the violence of the relativistic jetting-process
may break many nuclei into their  constituents.

In our admittedly sketchy model of the GRB engine, the time
structure of the $\gamma$-ray signal will be essentially that of the ejections. 
The energies of the GRB subpulses and the 
``waiting times'' between  them are nearly proportional
(e.g. van Paradijs et al. 1988). This is indicative of a process,
such as discontinuous
accretion, in which the ``reservoir'' must be replenished.


Let ``jet'' stand for the ensemble of CBs emitted in one
direction in a SN event. If a momentum imbalance between the 
opposite-direction jets is responsible for the large peculiar velocities
${\rm v_{NS}\approx 450\pm 90~ km~s^{-1}}$ (Lyne and Lorimer 1994) 
of neutron stars born in SNe, 
the jet kinetic energy 
must be, as we shall assume for our GRB engine, $\sim 10^{52}$ erg
(e.g. Dar and Plaga 1999).
For this choice 
the jet mass is relatively small:
${\rm M_{jet}\sim 1.5 \times 10^{-6}\,M_{NS}\,(10^3/\gamma)}$,
or ${\rm \sim 1.86\, M_\otimes}$ for the Lorentz factor 
$\gamma={\cal{O}}(10^3)$
that we independently estimate from the GRBs beaming angle,
from the average fluence per pulse observed by BATSE
in GRBs, and from the typical duration of these pulses.

The large peculiar velocities of pulsars may have 
been acquired well after their birth, in a ``second'' transition of
the cooled and spun-down
neutron star to a strange star or a quark star (Dar and De R\'ujula
2000b). In that case $\rm v_{NS}\!\sim\! 100$ km s$^{-1}$ ought to be used:
the observed velocity of young pulsars, millisecond pulsars
and stellar black holes in binaries and isolation (e.g. Toscano et al.
1999 and references therein). The typical jet energy would then
be $\sim\! 10^{51}$ erg and our results would have
to be accordingly scaled, but their dependence on $\rm E_{jet}$ is
weak and the conclusions are unaffected.

There are other events in which a variety of GRBs
could be produced by mechanisms similar to the ones we have
discussed: large mass accretion episodes in binaries including
a compact object, mergers of neutron stars with neutron stars
or black holes, the ``second transitions'' just discussed, etc.
In each case, the ejected cannonballs would make GRBs by
hitting stellar winds or envelopes, circumstellar mass or light.
We discuss only core-collapse SN explosions, as the GRBs 
they would produce
by our mechanism, although relatively ``standard'', satisfactorily
reproduce the general properties of the heterogeneous
ensemble of GRBs.
 
\section{The making of a GRB}

The making of a GRB is illustrated in Fig.~3.
In the time after core collapse ${\rm t_{fall}}={\cal{O}}(1)$ day,
as in Eq.~(\ref{tfall}), that it takes to 
``fuel and start'' the GRB engine, the SN shell  will have 
acquired a radius ${\rm R_s\sim 2.6\times 10^{14}}$ cm, 
for a typical velocity $\rm \sim c/10$ of the ejecta (we do not
scale all results with all the ratios of actual to typical
values, as we are only sketching the model).
If the CBs expand with  a velocity $\rm \beta_{exp}\, c$
in their rest system ($\rm \beta_{exp}\! =\! 1/\sqrt{3}$ for
a relativistic plasma), by the
time they hit the SN shell their transverse radius has grown to 
$\rm \beta_{exp}\, R_s / \gamma$,
or $\rm R_{CB}\sim 1.5\times 10^{11}\, cm$
for $\beta_{exp}=1/\sqrt{3}$ and $\gamma\sim 10^3$. 

The typical number of CBs per SN event we certainly cannot predict.
But the observed number of bursts in intense GRBs averages
to $\rm n_{CB}\!\sim\! 5$: the number ---in our model--- of well aimed and
relatively energetic (i.e. observable) GRBs per SN.
Let a CB have a fraction f of the total jet energy
and let ${\rm M_s}$ be the mass of the SN shell. The ratio of the
swept-up shell mass to the CB's energy is ${\rm M_s
c^2\,(R_{CB}/R_s)^2/(4\,f\, E_{jet})}$, that is ${\rm 6\times 10^{-4}/
f}$ for ${\rm M_s=10\, M_\odot}$ and ${\rm E_{jet}=10^{52}}$ erg. 
The ensemble of CBs (for which $\rm f=1$)
certainly blows a passage through
the SN shell, but, for f sufficiently small, the leading CBs are
significantly decelerated by the shell, opening up the possibility of CBs
colliding and merging after they exit the shell, which complicates the
nature and time structure of the GRB signal. The fraction of the SN shell
surface that is blown up by a CB is only
$\rm 5\times 10^{-7}\,(\beta_{exp}/0.1)^2\,(10^3/\gamma)^2$,
so that it appears unlikely that the CBs are aimed
precisely enough to hit the very same place: CB pile-up may be the
exception, not the rule.

The collisions of the CB nuclei ---whose energy is $\cal{O}$(1) TeV---
with those in the SN shell produce
hadronic cascades.  The SN shell and the CB are transparent to the neutrinos,
semitransparent to the muons and opaque to the photons constituting
these cascades. The CB inpact on the shell generates
 a very intense, narrowly
collimated and very short burst of sub-TeV neutrinos. The 
fraction of c.m. energy that
does not escape in neutrinos or muons ($\sim 1/3$) is converted into
internal black-body radiation of the CB.  The CBs become visible in
$\gamma$-rays when the obscuring column density of the shell approaches
one absorption length, ${\rm X_s(r)\sim X_\gamma\sim 10~g\, cm^{-2}}$,
resulting in a short burst of sub-TeV $\gamma$-rays from $\pi^0$ decays.
As a reheated CB expands, it becomes transparent and emits
its internal radiation, collimated and boosted by the CB's motion
and observed in the form of MeV $\gamma$-rays: one of the
pulses that constitute a GRB. Of all these issues (Dar and
De R\'ujula 2000a), we only discuss here the total energy in
GRBs and the time duration of their pulses. 

Approximately a third of the collision c.m. energy  is 
converted in the hadronic cascades into electromagnetic radiation (via 
$\pi^0$ production). Energy and momentum conservation
in the collision between a CB and the SN shell imply that
the internal heat acquired by the CB is ${\rm E_i\sim  
(\beta_{exp}/6\,\gamma)\,\sqrt{2\,f\,E_{jet}\, M_s\, c^2}}$.
The corresponding radiation pressure makes the CB expand
and cool at a rate ${\rm T_{CB}\sim 1/R_{CB}\sim 1/t}$,
until it becomes optically thin and releases its enclosed radiation.
For an ionized CB, this takes place when
${\rm R_{CB}\simeq [3\,M_{CB}\,\sigma_T/(4\, \pi\, m_p)]^{1/2}}$,
where $\rm \sigma_T\simeq 0.65\times 10^{-24}$ cm$^2$
is the Thomson cross section\footnote{Nuclei are disassembled
by the CB's collision with the SN shell, and neutrons decay
in $\sim\! 1$ s of observer's time, thus the number of electrons 
in the CB is $\rm M_{CB}/m_p$.}. 

The internal heat, reduced by expansion, 
is the energy the CB radiates
in its rest system, which is boosted and collimated by the CB's motion.
Local observers at angles $\theta$ to the CB's direction would
see this energy distributed as:
\begin{equation} 
{\rm {dE\over d\Omega} \simeq {\beta_{exp}^2\over \gamma^{3/2}}\,
\left[{2\,\gamma \over
(1+\gamma^2\, \theta^2)}\right]^3\,\left[{4\,\pi\, R_s^2\over
3\,\sigma_{_T}
}\right]^{1/2}\,{\sqrt{2\,m_p\,M_s}\, c^2\over 24\,\pi}}\,.
\label{funnel}
\end{equation} 
Remarkably, the CB's mass and energy have
dropped from the above expression,
except for the fact that, for the result to be correct, they must be
large enough for the CB to pierce the SN shell and remain relativistic.
For  a most likely viewing angle $\theta\!\sim\! 1/\gamma$, 
and $\gamma\! =\! 10^3$, one infers from Eq.~(\ref{funnel})
an isotropic energy  ${\rm E_{isot}
\simeq (\beta_{exp}/0.1)^2\, 6.4 \times 10^{53}\,erg}$. This energy ---or
$\rm n_{CB}$ times this energy in a GRB with that many pulses---
are in agreement with the range of observed isotropic
energies listed in Table I. The light from the SN shell is also
Compton up-scattered to MeV energies, but its contribution
to a GRB is subdominant relative to that of Eq.~(\ref{funnel})
(Dar and De R\'ujula 2000a).

The duration of a single CB-induced GRB pulse is the light
crossing time of the CB when it becomes transparent
to its radiation, i.e. ${\rm \sim R_{cb}/c}$ in its rest frame, or 
\begin{equation}
\rm t\sim \left[{3\,M_{CB}\,\sigma_T
\over 4\,\pi\, m_p}\right]^{1/2}
\; {1+\theta^2\gamma^2\over 2\,\gamma\,c}\,(1+z)\; .
\label{time}
\end{equation}
for the observer. 
For $z=1$, a most probable $\theta\!\sim\! 1/\gamma$, and our adopted values, 
$\rm E_{jet}=10^{52}$ erg and $\gamma=10^{3}$, 
$\rm t\sim 2.2\;f^{1/2}$ s, comparable to the
observed duration of GRB pulses. The total duration of a GRB is that
of the (unpredictable) firing-sequence time of CBs into the observer's 
viewing cone;
the duration of GRBs with only one observable pulse 
is given by Eq.~(\ref{time}).

The evidence for superluminal motion in GRBs, as we shall
argue, can only be measured in their afterglow phase.
The only case in which this can be done (SN1998bw/GRB 980425)
is already some 850 days old: it may be dimming out of
observability. Hence, we skip a more detailed 
discussion of the spectral properties of GRBs (Dar and De R\'ujula, 2000a)
to concentrate on the issue of superluminal motion of the
afterglow's source.

\section{GRB afterglows}

Far from their parent SNe, the CBs are slowed down by the interstellar
medium (ISM) they sweep, which has been previously ionized by the
forward-beamed CB radiation (travelling essentially at $\rm v=c$, the CB
is ``catching up'' with this radiation, so that the ISM has no time to
recombine). As in the jets and lobes of quasars, a fraction of the
swept-up ionized particles are ``Fermi accelerated'' to cosmic-ray
energies and confined to the CB by its turbulent magnetic field,
maintained by the same confined cosmic rays (Dar 1998b; Dar and Plaga
1999). The synchrotron emission from the accelerated electrons, boosted by
the relativistic bulk motion of the CB, produces afterglows in all bands
between radio and X-rays, collimated within an angle $\sim 1/\gamma(t)$,
that widens as the ISM decelerates the CB. When the CBs finally stop, they
enter a Sedov--Taylor phase, causing a temporal ``break'' in their
afterglow.

A CB of roughly constant cross section, moving in a
previously ionized ISM of roughly constant density, 
would lose momentum at a roughly constant rate, 
independent of whether the ISM
constituents are re-scattered isotropically
in the CB's rest frame, or their mass is added to that of the CB.
The pace of CB slowdown is 
${\rm d\gamma/dx=-\gamma^2/x_{0}}$, with ${\rm x_{0}=M_{CB}/
(\pi\, R_{CB}^2\, n\, m_p})$ and n the number density along
the CB trajectory. 
For $\gamma^2\gg 1$,
the relation between the length of
travel dx and the (red-shifted, relativistically aberrant) time of 
an observer at a small angle $\theta$ is 
${\rm dx=[2\, c\, \gamma^2/(1+\theta^2\,\gamma^2)]\,[dt/(1+z)]}$.
Inserting this into $\rm d\gamma/dx$ and integrating, we obtain:
\begin{equation}
{\rm {1+3\,\theta^2\gamma^2\over 3\,\gamma^3}=
{1+3\,\theta^2\gamma_0^2\over 3\,\gamma_0^3}+
{2\,c\, t\over (1+z)\, x_{0}}}\; ,
\label{gamoft}
\end{equation}
where $\gamma_0$ is the Lorentz factor of the CB as it exits
the SN shell.
The real root $\rm \gamma=\gamma(t)$ of the cubic Eq.~(\ref{gamoft})
 describes the CB slowdown with observer's time. 

Let the radiation emitted by a CB in its rest system
(bremsstrahlung, synchrotron, Compton-boosted 
self-synchrotron) be isotropic and have an energy flux density
$\rm F_0(\nu_0)$. This radiation is boosted and collimated 
by the CB's motion, and its time-dependence is
modified by   the observer's 
time flowing  $(1+z)\,(1+\gamma^2\theta^2)/(2\gamma)$ times
faster than in the CB's rest system. 
For $\gamma \gg 1$,
an observer at small $\theta$ sees an energy flux density:
\begin{eqnarray}
{\rm F[\nu] \simeq \left({R_{CB}\over D_L}
\right)^2\;}&&{\rm \left[{2\gamma\over 
1+\gamma^2\theta^2}\right]^3\; }\, \times\nonumber \\
&&{\rm F_0\left(\nu\,[1+z]\;{1+\gamma^2\theta^2
\over 2\,\gamma}\right)\; A(\nu,z)
}\, ,
\label{sync}
\end{eqnarray}
    
with $\rm \gamma=\gamma(t)$ as in Eq.~(\ref{gamoft}) and
$\rm A(\nu,z)$ an eventual absorption dimming.

During the afterglow regime, the CBs emission is dominated by electron
synchrotron radiation from the magnetic field in the CB.  At frequencies
above the plasma frequency, a CB becomes transparent to this radiation
when the attenuation length due to electron free--free transitions exceeds
its radius.  The spectral shape is ${\rm F_0\sim \nu_0^{-\alpha}}$, with
${\rm \alpha=(p-1)/2}$ and p the spectral index of the electrons. For
equilibrium between Fermi acceleration and synchrotron and Compton
cooling, $\rm p \approx 3.2$ and $\alpha\approx 1.1$, while for small
cooling rates, $\rm p \approx 2.2$ and $\alpha \approx 0.6$ (Dar and De
R\'ujula 2000c), or $p\simeq1.2$ and $\alpha\simeq 0.1$ if Coulomb 
losses dominate. 
At very low radio frequencies self-absorption becomes important and
$\alpha \approx -1/3~(2.1)$ for optically thin (thick) CBs.
For a detailed modelling of synchrotron radiation from
quasar lobes, see, for instance, Meisenheimer et al. (1989).

Since $\rm \gamma(t)$, as in Eq.~(\ref{gamoft}), is a decreasing
function of time, the afterglow described by Eq.~(\ref{sync})
may have a very interesting behaviour.
An observer may initially be outside the beaming cone: 
$\theta^2\gamma^2\! >\! 1$, as we shall argue to be the case for
GRB 980425, for which we estimate $\gamma_0^2\,\theta^2\!\sim\! 100$ 
(other relatively dim GRBs in Table I, such as 970228 and 970508, 
may also be of this type).
The observed afterglow would then initially rise with time. 
As $\gamma$ decreases, the cone
broadens, and at $\gamma\theta\!\sim\! 1$ the time dependence changes  
towards ${\rm  F_\nu\propto t^{-(3+\alpha)/3}\sim t^{-1.4}}$,
while the $\nu$ dependence stays put at $\nu^{-\alpha}$.
Such a behaviour may explain the puzzling initial rise of the
optical afterglow of GRBs 970228 and 970508,
as well as the second peak around $\rm t_p\sim 3$--$\rm 4\times 10^6\, s$ 
in the unresolved radio emission 
from SN1998bw/GRB 980425  (Kulkarni et al. 1998a; Frail et al. 1999),
if it corresponds to the GRBs afterglow.

The afterglow of GRB 970508 is shown in Fig.~2, and compared with
our prediction Eqs.~(\ref{gamoft},\ref{sync}), for the measured index
$\alpha=1.1$. The adjusted parameters are
the height and time of the afterglow's peak 
and $\theta=10/\gamma_0$.
The figure is for a single CB; with a few of them at chosen times
and relative fluxes,
it would be easy to explain the early ``warning shots'' at $\rm t<1$ day
and the abrupt rise at $\rm t=1$ to 2 days. At $\rm t\gg t_p$, however,
they would add up to a single curve like the one shown in the figure.

As a CB is finally stopped by the ISM, it enters a Sedov--Taylor phase.
The CB's radius increases as ${\rm t^{2/5}}.$
The Lorentz factor of electrons decreases like ${\rm  t^{-6/5}}$.
In equipartition, the magnetic field decreases like ${\rm t^{-3/5}}$.  
Wijers et al. (1997) have shown that these facts lead to an afterglow
decline ${\rm F_\nu\sim t^{-(15\alpha-3)/5}\sim  t^{-2.7}}$
for  $\alpha\sim 1.1$,   
as observed for the late-time afterglows of
microquasars and of some GRBs.

\section{GRB 980425: a special case}

In the list of Table I, GRB 980425 stands out in two ---apparently 
contradictory---
ways: it is, by far, the closest ($\rm z=0.0085$, $\rm D_L=39$ Mpc) 
and it has, by far, the smallest implied spherical energy: ${\rm 
8.1\times 10^{47}}$ erg,  4 to 6 orders of magnitude smaller than that 
of other GRBs. If GRB 980425 was not abnormal, Eq.~(\ref{funnel}) 
tells us that it must have been beamed  at an angle
$\theta\approx 10/\gamma_0$ relative to our line of sight. 
The probability of a jetted GRB to be pointing to us
within an angle $\theta\leq 10/\gamma_0 \sim 10^{-2}$  is
${\rm P\sim \pi\theta^2/(4\pi)\approx 2.5\times 10^{-5}}$. 
For this GRB, BATSE observed a single wide peak of $\sim 5$ s 
duration and a maximum
flux ${\rm (3\pm 0.3)\times 10^{-7}~erg \,cm^{-2}\,s^{-1}}$ (Soffita et
al. 1998). With the BATSE sensitivity of ${\rm  6\times 10^{-2}
\,cm^{-2}\,s^{-1}}$ in the range ${\rm 50<E_\gamma<300~keV}$, such a burst 
can be detected up to $\sim 150$ Mpc.  For our choice of $\rm H_0$, the local
luminosity density is ${\rm \sim 1.3 \times 10^{-2}\,  L_*\, Mpc^{-3}}$
(e.g. Ellis 1997),  yielding ${\rm \sim 3.7\times 10^3} $ SNe a year 
within 150 Mpc. Thus, the mean time to observe a SN--GRB association 
with $\theta>10/\gamma_0$ is $\sim$ 10 y, consistent with a single 
detection, since the launch of  BATSE, of a GRB with the odd 
properties of 980425.

\section{Is the superluminal motion of GRB 980425 observable?} 

As we have discussed, the peculiarities of of GRB 980425 can be understood
if its source was ``fired''  from SN1998bw with a bulk-motion Lorentz factor
$\gamma \sim 1000$, at an angle $\theta\approx 10/\gamma$ relative to our
line of sight.  Its transverse superluminal
displacement, $\rm D_T$, from the SN position can be obtained by 
time-integrating
$\rm v_T$, as in Eq.~(\ref{superlum}), using $\rm \gamma(t)$ as in
Eq.~(\ref{gamoft}). The result can be reproduced, to better than 10\%
accuracy, by using the approximation $\rm v_T\sim 2 \gamma^2\,\theta\,c$,
valid for $\gamma\! <\! 1/\theta$ (or $\rm t\! >\! t_p$, the afterglow's
peaktime):  
\begin{equation} 
{\rm D_T\simeq{2\,c\, t_p\over \theta} \left[{t\over t_p}\right]^{1/3}}.  
\end{equation} 
At present ($\rm t\!\sim\! 850\, d$),
the displacement of the GRB from the initial SN/GRB position is $\rm
D_T\!\sim\! 20\,(\gamma_0/10^3)$ pc, corresponding to an angular
displacement $\Delta\alpha\!\sim\!100\,(\gamma_0/10^3)$ mas, for $\rm
z=0.0085$, $\rm t_p\!\sim\! 3.5\times 10^6$ s.
 
The two sources, a few tens of mas away, may still be resolved by HST. 
In Fig.~4 we show our prediction for the late-time V-band light curve of
SN1998bw/GRB 980425. The SN curve is a  fit by Sollerman 
et al. (2000) for energy deposition by $\rm ^{60}Co$ decay  
in an optically thin SN shell.  The GRB light curve is our predicted 
afterglow for GRB 9980425, as given by Eqs.~(\ref{gamoft},\ref{sync}),
and constrained to peak at the position of the second peak in
the radio observations (Kulkarni et al. 1998a; Frail et al. 1999).
 The fitted normalization  
is approximately that of the mean  GRB afterglow of the GRBs in Table I
suppressed by the same factor as its $\gamma$-ray fluence relative to 
the mean $\gamma$-ray fluence in Table I.  
The joint system has at present (day $\sim\! 850$) an extrapolated   
magnitude $\rm V\sim 26$ (Fynbo et al. 2000). 
It  can still be resolved  from its
host galaxy ESO-184-G82 by HST and, perhaps, by VLT in good seeing
conditions.  An extrapolation of the V-band late time
curve of SNR1998bw (Sollerman et al. 2000) suggests
that the present magnitude of the SN is $\rm V\sim 28$ , which is near
the detection limit of HST and is dimming  
much faster than $\rm ^{60}Co$-decay would imply.

In the GHz radio band, the system has been last observed  in February 1999
by ATCA (Australia Telescope Compact Array) to have $\sim$~mJy
flux density (Frail et al. 2000). It approached a power-law time decline
with a power-law index -1.47. If a Sedov-Taylor break in the radio afterglow 
of GRB 980425 has not occurred yet, its
spectral density may still be strong enough to determine its position with
VLA and VLBI to better than mas precision. If the second peak is the GRB's
afterglow, the two radio centroids should now be separated by
$\sim\!\gamma_0/10$ mas (or by $\sim\!\gamma_0/20$ mas, if the first peak
is the afterglow).  A refined location from a reanalysis of the early ATCA
observations (Kulkarni et al 1998a; Frail et al. 1999) of the initial SN
and of the late afterglow may also reveal a superluminal displacement.

If the GRB  afterglow  has entered the late
fast-decline phase seen in some GRB afterglows and in quasar and
microquasar ``afterglows'' from jetted ejections (observed power-law index
$-2.7 \pm 0.3$),  a further delay in follow-up observations can make it
very difficult or impossible to detect and resolve the GRB/SNR radio image
into its two  predicted images.  

A GRB as close as GRB 980425 (z = 0.0085)  should occur only once every 
$\sim\!10$ years. For typical GRBs ($\rm z\!\sim\! 1$) there is no hope of 
resolving them with HST into two separate SN and GRB images. Resolving
them with VLBI would also be arduous. For these reasons, we
exhort interested observers to consider immediate high-resolution optical
(STIS)  and radio (VLA and VLBI) follow-up observations of SN1998bw and
the afterglow of GRB980425. 

\section{Conclusions}

We have argued that GRBs and their afterglows may be produced
by the firing of extremely relativistic cannonballs in SN explosions.
For the GRB to be observable, the CBs must be close to
the line of sight, implying that their afterglows would appear to
move superluminally. Only one of the located GRBs (980425)
is close enough to us for this superluminal displacement to be observable
with the currently available resolution. Its afterglow may by
now have reached a ``break'' and be too dim to be seen. Or it may not.  
If observed, the superluminal displacement of this GRB's afterglow would
be a decisive card in favour of cannonballs, as opposed to stationary
fireballs.

{}

\newpage
\vskip 0.3 true cm

{\bf 
\noindent
Table I - Gamma ray bursts of known redshift z}
\begin{table}[h]
\hspace{-.5cm} 
\begin{tabular}{|l|c|c|c|c|c|c|c|c|}
\hline
\hline
GRB   &z &Ref  &D$_{\rm L}$$^a$ &${\rm F_\gamma}$$^b$
&${\rm E_\gamma}^c$ & M$\; ^d$  &Ref.\\
\hline
970228   &0.695  &1   &4.55  &0.17  & 0.025     & 25.2       &15 \\
970508   &0.835  &2   &5.70  &0.31  & 0.066     & 25.7       &16 \\
970828   &0.957  &3   &6.74  &7.4    & 2.06       &  ---         &17 \\
971214   &3.418  &4   &32.0  &1.1    & 3.06       & 25.6       &18 \\
980425  &0.0085 &5   &0.039&0.44  & 8.14 E-6& 14.3       &19 \\
980613   &1.096  &6   &7.98  &0.17  & 0.061     & 24.5       &20 \\
980703   &0.966  &7   &6.82  &3.7    & 1.05       & 22.8       &21 \\
990123   &1.600  &8   &12.7  &26.5  & 19.8       & 24.4       &22 \\
990510   &1.619  &9   &12.9  &2.3    & 1.75       & {\it 28.5}       &23 \\ 
990712   &0.430  &10 &2.55  & ---    & ---           & 21.8       &24 \\
991208   &0.706  &11 &4.64  &10.0  & 1.51       &$>$ 25      &25 \\
991216   &1.020  &12 &7.30  &25.6  & 8.07       & 24.5       &26 \\
000301c &2.040  &13 &17.2  &2.0    & 2.32       & 27.8       &27 \\
000418   &1.119  &14 &8.18  &1.3    & 0.49       & 23.9       &28 \\

\hline
\hline
\end{tabular}
\end{table}
\vskip -0.3 true cm
\noindent
{\bf Comments:} $a$: Luminosity distance in Gpc (for $\rm \Omega_m=0.3,
\; \Omega_\Lambda=0.7$ and ${\rm H_0=65\, km\, s^{-1}\,Mpc^{-1}}$.
$b$: BATSE $\gamma$--ray fluences in units of
$10^{-5}$ erg cm$^{-2}$. $c$: (Spherical) energy in units of  $10^{53}$ ergs.
$d$: R-magnitude of the host galaxy, except for GRB 990510, for
which the V-magnitude is given.
\noindent
{\bf References}:
1: Djorgovski et al. 1999;
2: Metzger et al. 1997;
3: Djorgovski et al. 2000;
4: Kulkarni et al. 1998b;
5: Tinney et al. 1998;
6: Djorgovski et al. 1998a;
7: Djorgovski et al. 1998b;
8: Kelson et al. 1999;
9: Vreeswijk et al. 1999a;
10: Galama et al. 1999;
11: Dodonov et al. 1999;
12: Vreeswijk et al. 1999b;
13: Feng et al. 2000;
14: Bloom et al. 2000;    
15: Fruchter et al. 1999b;
16: Bloom et al. 1998a;
17: Djorgovski et al. 2000a;
18: Odewahn et al. 1999;
19: Galama et al. 1998;
20: Djorgovski et al. 2000b; 
21: Bloom et al. 1998b;
22: Fruchter et al. 1999c;
23: Fruchter et al. 1999d;
24: Hjorth et al. 1999;
25: Diercks et al. 2000;
26: Djorgovski et al. 1999;
27: Fruchter et al. 2000;
28: Metzger et al. 2000.

\newpage

\begin{figure}[t]
\begin{tabular}{cc}
\hskip 2truecm
\vspace*{2cm}
\hspace*{-1.7cm}
\epsfig{file=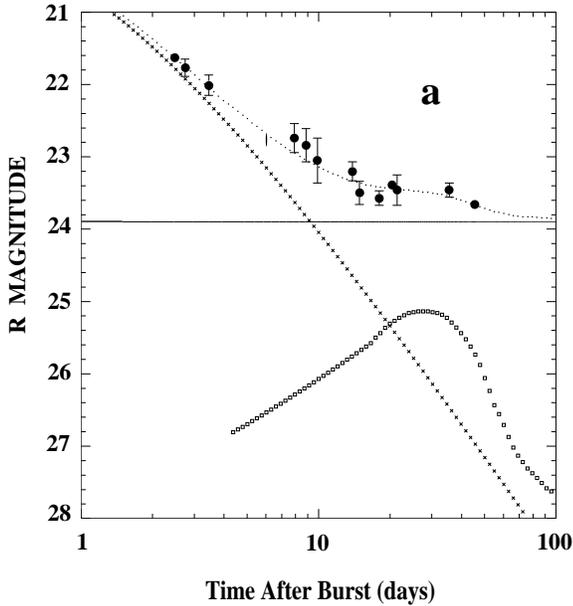,width=7.5cm} \\
\hspace*{.5cm}
\epsfig{file=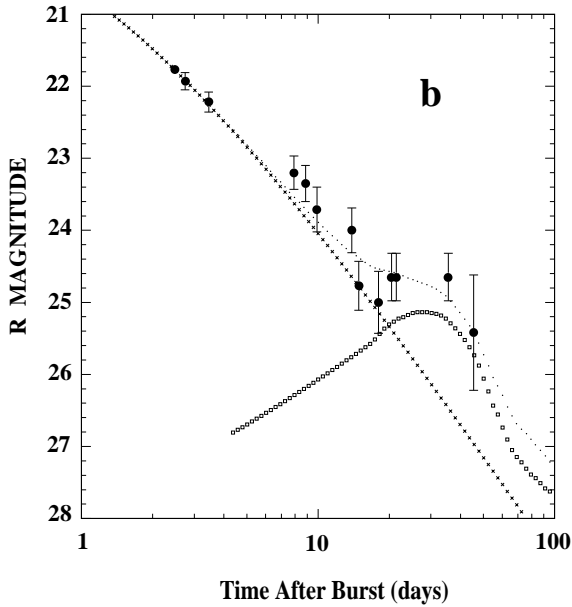,width=7.5cm}
\end{tabular}
\caption{Comparisons between our predicted R-band light curve 
for GRB 000418 (dotted lines) and the observations as compiled 
by Klose et al. (2000). a) Without subtraction of the host
galaxy's contribution: the straight line with $\rm R=23.9$ (Fruchter et al.
2000). b) With the host galaxy subtracted.
The CB's afterglow is given by Eq.(\ref{gamoft},\ref{sync})
with spectral index $\alpha=1.9$  (Klose et al. 2000) and 
is indicated by crosses. The contribution 
from a SN1998bw-like SN placed at z=1.11854, as in
Eq.(\ref{bw}) with 
Galactic extinction ${\rm A_R=0.09}$ magnitudes,
is indicated by open squares.
The dotted line is the sum of contributions. The SN bump is
clearly discernible.}
\end{figure}

\begin{figure}
\begin{center}
\vspace*{1cm}
\hspace*{-.7cm}
\epsfig{file=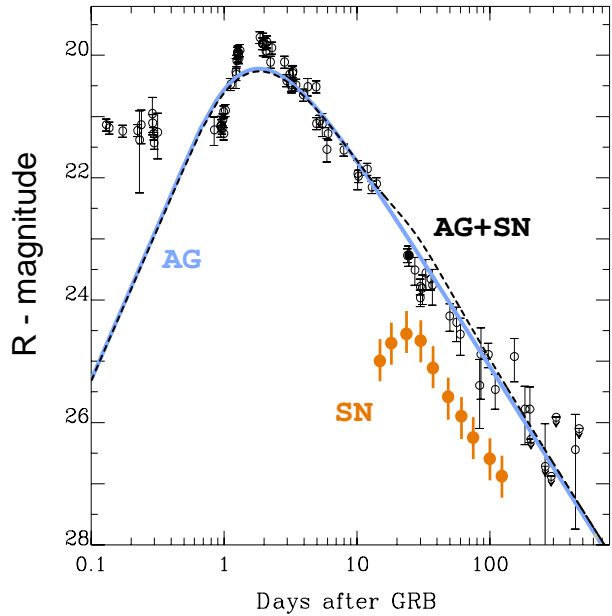,width=8cm}
\caption{The R-band light curve of the afterglow of GRB 970508 as
compiled by 
Fruchter et al. (1999a) 
with a constant (R = 25.2 magnitude) host galaxy  
subtracted from all the measurements. The blue ``AG'' curve is 
given by Eqs.~(\ref{gamoft},\ref{sync}). 
The contribution from a SN1998bw-like SN,
placed at the GRB redshift $\rm z=0.835$, given by Eq.(\ref{bw}),
is indicated
(in red) and makes very little difference when added to the afterglow.}
\vspace*{-0.5cm}
\end{center}  
\end{figure}

\begin{figure}
\begin{center}
\vspace*{.003cm}
\hspace*{-0cm}
\epsfig{file=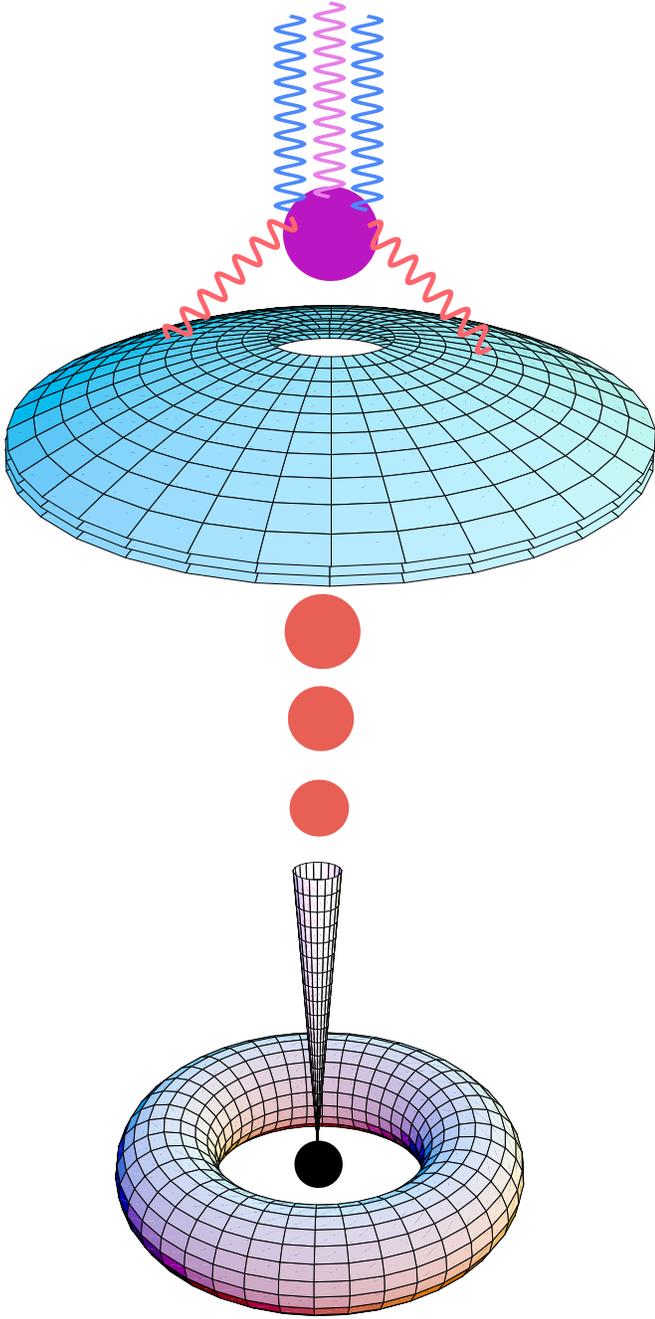,width=8.7cm}
\caption{An ``artist's view'' (not to scale) of the CB model
of GRBs and their afterglows. A core-collapse SN results in
a compact object and a fast-rotating torus of non-ejected
fallen-back material. Matter accreting (and not shown)
into the central object produces
a narrowly collimated beam of CBs, of which only some of
the ``northern'' ones are depicted. As these CBs pierce the SN shell,
they heat and reemit photons. They also scatter light from the shell.
Both emissions are Lorentz-boosted and collimated by the CBs' relativistic motion.}
\vspace*{-0.5cm}
\end{center}  
\end{figure}

\begin{figure}
\begin{center}
\vspace*{1cm}
\hspace*{-.7cm}
\epsfig{file=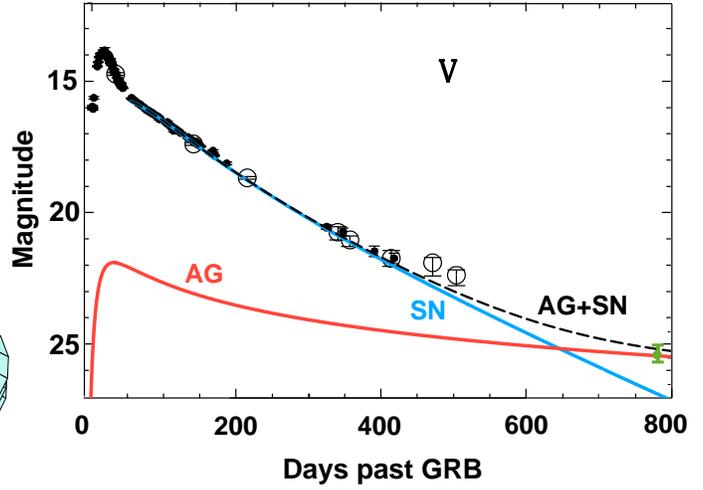,width=9cm}
\caption{The V-band light curve of 
SN1998bw/GRB 980425, with   
the blue ``SN'' curve a fit to the SN by Sollerman et al.~(2000).
 The red ``AG'' curve is our predicted 
afterglow, as given by Eqs.~(\ref{gamoft},\ref{sync}), fit
to peak at the position of the observed second radio peak, and
to reproduce the most recent observation at $\rm d=778$.
The dashed curve is the total.}
\vspace*{-0.5cm}
\end{center}  
\end{figure}

\end{document}